\documentclass[a4paper]{article}
\usepackage{graphicx}
\usepackage{float}
\usepackage{amsmath}
\usepackage{amssymb}
\usepackage{bm}
\usepackage{times}
\usepackage{a4wide}

\newcommand{\rmd}{{\rm d}}
\newcommand{\rme}{{\rm e}}
\newcommand{\rmi}{{\rm i}}

\usepackage[affil-it,auth-sc]{authblk}
\title{Half integer features in the quantum Hall Effect:\\ experiment and theory}

\author[1,2]{Tobias Kramer}
\author[2,3]{Eric J.\ Heller}
\author[4]{Robert E.\ Parrott}
\author[5]{Chi-Te Liang}
\author[6]{C.~F.\ Huang}
\author[5]{Kuang Yao Chen}
\author[7]{Li-Hung Lin}
\author[8]{Jau-Yang Wu}
\author[8]{Sheng-Di Lin}

\affil[1]{Institute for Theoretical Physics, University of Regensburg, 93040 Regensburg, Germany}
\affil[2]{Department of Physics, Harvard University, Cambridge, MA~02138, USA}
\affil[3]{Department of Chemistry and Chemical Biology, Harvard University, Cambridge, MA~02138, USA}
\affil[4]{School of Engineering and Applied Science, Harvard University, Cambridge, MA~02138, USA}
\affil[5]{Department of Physics, National Taiwan University, Taipei, Taiwan 106, R.O.C.}
\affil[6]{National Measurement Laboratory, Center for Measurement Standards, Industrial Technology Research Institute, Hsinchu, 
Taiwan 300, R.O.C.}
\affil[7]{Department of Applied Physics, National Chiayi University, Chiayi, Taiwan 600, R.O.C.}
\affil[8]{Department of Electronics Engineering, National Chiao Tung University, Hsinchu, Taiwan 300, R.O.C.}

\begin{document}

\maketitle

\begin{abstract}
The quantum Hall  effect is one of the most important developments in condensed matter physics of the 20th century. 
The standard explanations of the famous integer quantized Hall plateaus  in the transverse resistivity are   qualitative,  and involve  assumptions about disorder, localized states, extended states,  edge states,   Fermi levels  pinned by Landau levels, etc.
These standard narratives give  plausible reasons for the existence of the plateaus, but  provide little in the way of even a qualitative understanding of  the shape and width of the Hall plateaus,  much less a first principles calculation \cite{Mahan2000a,Davies1998a}.
The injection model  presented in this paper changes that situation.   Rather than focusing on the middle of the Hall device, we follow the electrons to their source: one corner of the Hall bar and its steep electric field gradients.   We find the   entire resistivity  curve including the Hall plateaus  is calculable as a function of magnetic field, temperature, and current.   The new higher current experiments  reported here show remarkable half integer features  for the first time; these are faithfully reproduced by the injection theory.
The Hall plateaus and half integer inflections are shown to result from  the local density of states appropriate to the magnetic field and  the strong electric field gradient at the injection corner.
\end{abstract}

The classical Hall effect generates a highly non-trivial electron flow pattern, electric field, and potential distribution, which originate from the boundary conditions imposed by the device geometry and the Lorentz force acting on moving charges \cite{Beer1963a}. Extremely high electric fields are present in the vicinity of two opposite corners of a Hall device (Figure 1). Experiments have confirmed that this picture prevails in the quantum Hall regime  and that electrons enter the device in this high-field zone, where dissipation and heating take place \cite{Knott1995a,Ahlswede2001a,Kawano1999a,Ikushima2007a}.
Nonetheless the  corners and their high-field injection zones do not play a significant role in any of the standard narratives about the integer quantum Hall effect (IQHE). Discussions focus  instead on the center section of the Hall bar, including the edges.   The de-emphasis of contacts and the injection point  goes so far as  introduction of a fictitious translational invariance,  crucial  for example to Laughlin's   elegant gauge argument for the  integer  plateaus \cite{Laughlin1981a}.    However   the gritty job of understanding   the flux of electrons through the corner from first principles has as its reward a calculable, quantitative understanding of    the transverse  resistivity  Hall plateaus.      A satisfying  continuity would prevail with  the theory of quantum point contacts (QPCs)  and two-dimensional electron gases (2DEGs) in the absence of a magnetic field, in which conductance is  calculable in terms of  the  injection of electron flux  at the QPC,  without having to worry about  the fate of the electrons after they leave the QPC,   apart from any backscattering.  Because of the  presence of a strong magnetic field  leading electrons along potential contours and along edges,     backscattering   is not an issue  in quantum  Hall experiments. 

\begin{figure}[t]
\begin{center}
\includegraphics{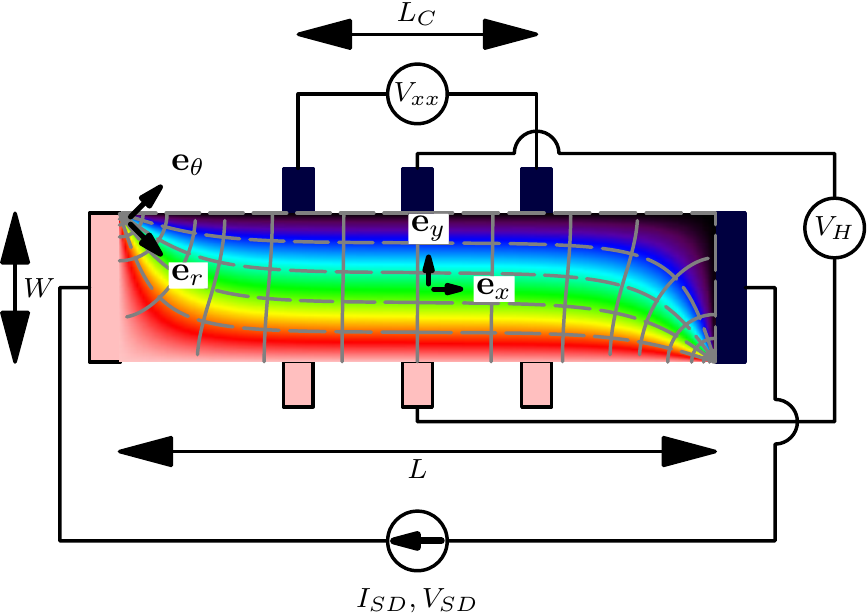}
\end{center}
\caption{
\textbf{Schematic picture of the Hall bar with attached voltage probes.}
The potential (dashed gray line) and electric field (solid gray line) distribution in a Hall bar. The color range from red to blue denotes the value of the potential. The equidistant equipotential lines at the center of the device follow the cartesian unit vector $\mathbf{e}_x$, while near the corner the electric field strength increases (hot spot) and the equipotential lines follow the radial unit vector $\mathbf{e}_r$ and run isotropic perpendicular to quarter circles. The source-drain voltage $V_{SD}$ is experimentally adjusted to deliver a constant direct current $I_{SD}$, and the Hall resistivity is obtained by measuring the ratio $\rho_{xy}=\frac{V_H}{I_{SD}}$, while the longitudinal resistivity is obtained via $\rho_{xx}=\frac{V_{xx}}{I_{SD}}\frac{W}{L_C}$.
\label{fig:conformal_map}
}
\end{figure}

The key ingredients to such an injection theory are knowing the local density of states at the injection point, and appropriate convolution with the finite temperature  Fermi distribution.  At the  injection corner of the Hall device,  steep and rapidly changing electric field gradients exist.  This would seem to make calculations very difficult, especially considering the strong magnetic field which is present.  Time-dependent wave packets  propagated by fast Fourier transform  have proven  to be extremely useful  tools for understanding and calculating quantum point contact  injection physics, including thermal averaging,  impurity  scattering,  small angle scattering due to donor atom density fluctuations, the quantum point contact potential, etc. \cite{Topinka2000a, Topinka2001a, Aidala2007a}.  In the  harsh environment of the corner, a novel wavepacket approach developed  earlier \cite{Kramer2008a}  allows calculation of quantum electronic transport within the stringent boundary conditions imposed by the device geometry.

The sample studied in this work is a modulation-doped GaAs/AlGaAs heterostructure grown by molecular beam epitaxy (MBE). The following layer sequence is grown on a GaAs semi-insulating substrate: 1~$\mu$m GaAs, 20~nm undoped Al$_{0.33}$Ga$_{0.67}$As, 40~nm Si-doped Al$_{0.33}$Ga$_{0.67}$As, and 10 nm GaAs cap layer. The device was made into a Hall bar pattern by standard optical lithography and etching processes and a Ti/Au gate was evaporated on the surface. The width of the Hall bar and the center-to-center distance between the two voltage probes  used to measure the longitudinal resistance are 80~$\mu$m and 720~$\mu$m,  respectively. At gate voltage $V_g =0$, the carrier density of our 2DEG is $\approx 2.27 \times 10^{15}$~m$^{-2}$ with a classical mobility of $\mu \approx 9.4$~m$^{2}/$Vs at $T \approx 0.3$~K. Four-terminal longitudinal resistivity $\rho_{xx}$ and Hall resistivity $\rho_{xy}$ were measured in a top-loading He$^{3}$ cryostat using standard ac phase-sensitive lock-in techniques at a frequency of 87 Hz. In our studies, current-dependent resistivity measurements were performed at a fixed lattice temperature of $T \approx 0.3$~K and $T$-dependent resistivity measurements were performed at a fixed current of 100~nA.

\begin{figure}[t]
\includegraphics[scale=0.85]{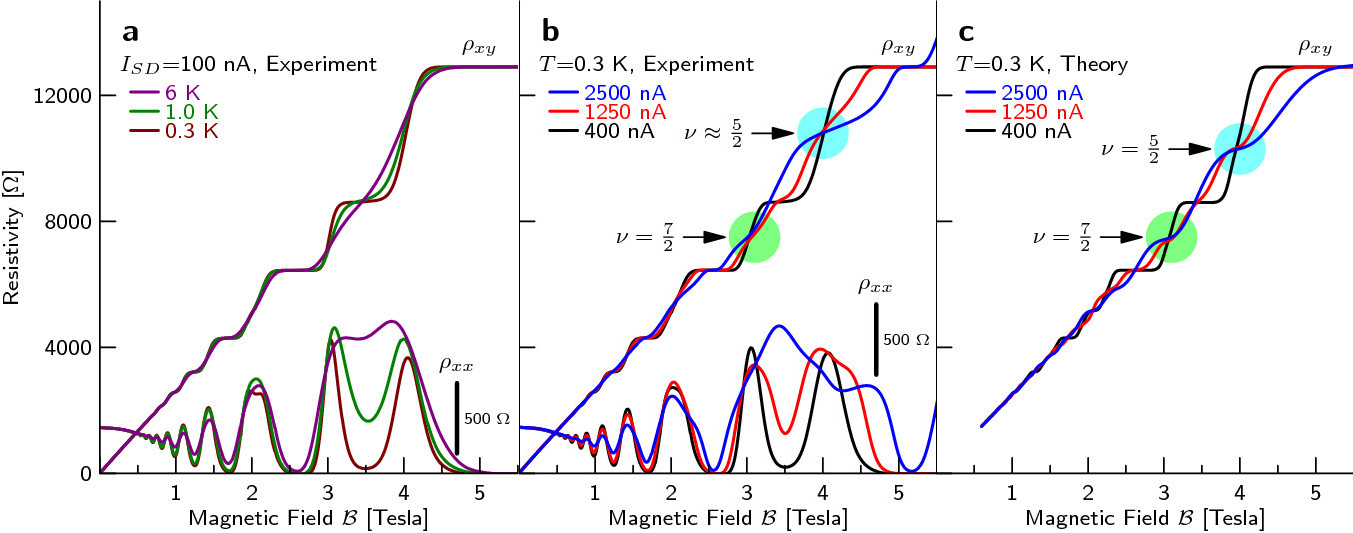}
\caption{
\textbf{Hall resistivity $\rho_{xy}$ as a function of the magnetic field.}
\textbf{a}, 
Experimental Hall longitudinal resistivity $\rho_{xx}$ and Hall resistivity $\rho_{xy}$ as a function of the magnetic field for a constant current and a series of increasing temperatures.
\textbf{b}, 
Experimental Hall resistivity $\rho_{xy}$ as a function of the magnetic field for a constant lattice temperature and a series of increasing currents. Intersection points with decreasing slope exist at filling factor $\nu=\frac{7}{2}$ and slightly above filling factor $\nu=\frac{5}{2}$.
\textbf{c},
Theoretical calculation of the Hall resistivity $\rho_{xy}$ for increasing currents.
The intersection points at filling factors $\nu=\frac{5}{2},\frac{7}{2}$ are prominently visible. 
\label{fig:rho_xy}}
\end{figure}

Figure~2a shows the measured $\rho_{xx}$ and $\rho_{xy}$ for the Hall device for a fixed current and a series of increasing temperatures, and Figure~2b for a fixed temperature and a series of increasing currents. While in both series a shrinking of the integer plateaus is visible, the increasing current series evolves very differently from the increasing temperature series. Both series share invariant intersection points around the center of integer plateaus. The current series however displays another series of intersection points, one very close to $\rho_{xy}=\frac{2 h}{7 e^2}$ (corresponding to the filling factor $\nu=\frac{7}{2}$), and one in the vicinity of $\nu=\frac{5}{2}$. These intersection points are not present in the fixed-current, varying temperature series. At these half-integer filling factors the slope of $\rho_{xy}({\cal B})$ decreases with increasing current. This is best analyzed in terms of the derivative of the Hall resistivity $\partial\rho_{xy}/\partial{\cal B}$, which provides a measure of the local density of states (LDOS) at the injection region. In Fig.~3b, the current induced gaps at filling factors  $\nu=\frac{5}{2}$ and $\frac{7}{2}$ are clearly visible as local minima (marked by arrows) of $\partial\rho_{xy}/\partial{\cal B}$, whereas the constant current series in Fig.~3a shows local maxima at the same points, which get broadened due to an increasing $k_B T$, where $k_B$ is the Boltzmann constant. The striking difference between the appearance of the local maxima in the temperature series and the local minima in the current series is not explained within conventional IQHE theories, where the effect of a higher current is interpreted as a change in temperature, which merely broadens Landau levels without inducing the formation and widening of a gap in the center of a Landau level at half-integer filling factors.

\begin{figure}[t]
\includegraphics[scale=0.85]{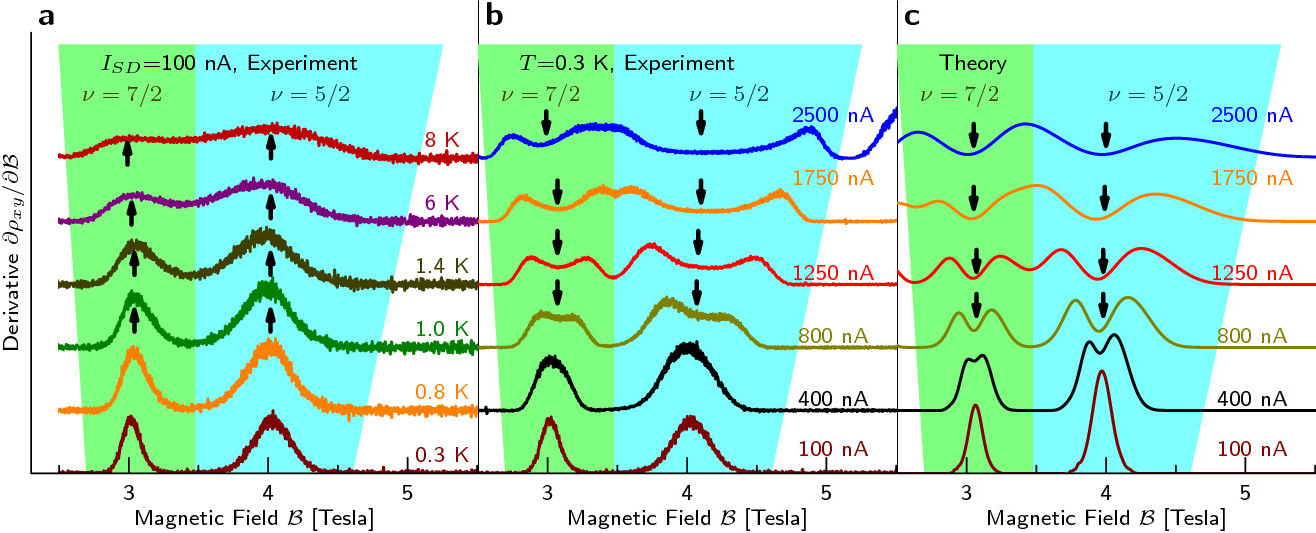}
\caption{
\textbf{Derivative of the Hall resistivity $\partial \rho_{xy}/{\partial \cal B}$.}
\textbf{a}, Constant current, series of increasing lattice temperatures. The up-arrows mark local maxima at filling factors $\frac{7}{2}$ and $\frac{5}{2}$.
\textbf{b}, Constant lattice temperature, series of increasing currents. In the increasing current series the maxima at ${\cal B}=3$~Tesla and $4$~Tesla turn into local minima, in contrast to the temperature series. These minima (marked by down-arrows) within the centers of the spin-split Landau level $n=1$ at $\nu=\frac{5}{2}$ and $\nu=\frac{7}{2}$ are imprinted by the varying substructure of the Landau levels in the hot-spot region of the device.
\textbf{c}, Calculated derivative of the Hall resistivity $\partial \rho_{xy}/{\partial \cal B}$ as a function of the magnetic field. 
}
\end{figure}

The usual picture of the shape of the density of states (DOS) as a function of energy is the following:
In the absence of disorder, the DOS in a magnetic field shows $\delta$-peaks at the Landau energies, which are assumed to be broadenend due to the presence of disorder and also to split into an extended band centered in the middle of each Landau level with associated states connecting both ends of the Hall bar, and a bordering localized state band, where no transport can occur.
We believe this central argument in traditional discussions is in fact not the key to an understanding of the IQHE.  It is much more important to understand the {\it local}  density of states at the point where electrons enter the device; this is the ``injection theory'' idea. 
Our model of the IQHE incorporates some aspects of the traditional narrative, namely that many-body effects can be incorporated at the mean-field level and an effective mass description of the electron prevails. Apart from these assumptions, a complete theory of the IQHE should incorporate all the known features of the experiment, unless they can be proven to be absolutely unimportant.
These features must include the finite geometry of the Hall bar, the presence of a non-vanishing current flow between the singular potential at two opposite corners, and the random background potential within the device.

\begin{figure}[t]
\begin{center}
\includegraphics[scale=1.0]{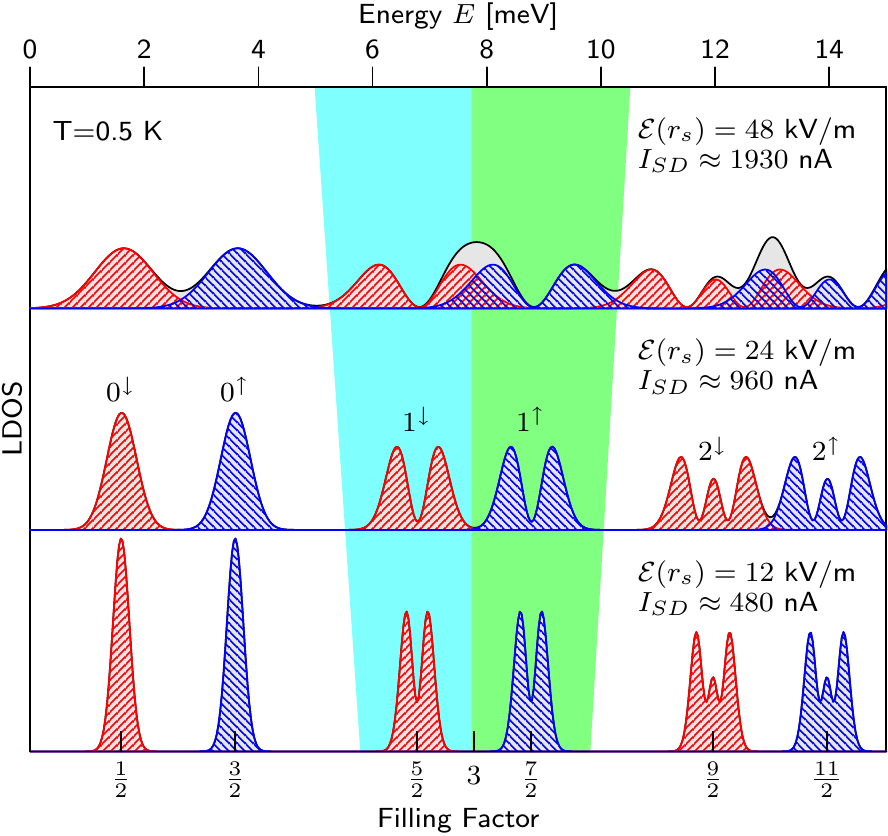}
\end{center}
\caption{
\textbf{Theoretical calculation of the LDOS}
(with spin splitting, $g^{*}=-0.77$) for the magnetic field ${\cal B}=3$~Tesla for different electric field strengths at a fixed distance of $220$~nm from the singularity. Note that the zeroes at the filling factors $\frac{7}{2}$ and $\frac{5}{2}$ are stable against variations of the electric field strength, whereas the shoulders start to overlap and turn the $\nu=3$ integer gap into a region of an enhanced LDOS.
}
\end{figure}

Near the corner, the local electric field is amplified 300 times compared to its value in the center of the Hall bar. Since dissipation occurs  precisely in this region and current is emitted into the device through Ohmic contacts, which connect the macroscopic part of the device with the 2DEG, it is crucial to derive the LDOS there. For the experimentally fixed current, we solve the self-consistent relation between the voltage drop across the corner (which broadens and modulates the LDOS) and the quantum mechanical current (see the Appendix).
The LDOS in a quantum Hall system  (Figure~4) shows a very different behaviour from the DOS of a disordered system: the LDOS drops to zero in the middle of even Landau levels, and thus transport is suppressed around the center of these Landau levels, as predicted before in a uniform electric field environment \cite{Kramer2005c,Kramer2003a,Kramer2003d,Kramer2006a}. 

The gaps in the LDOS disappear at low currents due to the convolution with the temperature broadened Fermi-Dirac distribution. It is remarkable that $\rho_{xy}$ is not affected by local disorder perturbations in the injection region, since the corner geometry enforces a much stronger electric field gradient than the gradient caused by disorder fluctuations.  The  Ohmic resistivity $\rho_{xx}$  is probed later in the device, where the amplification effect of the Hall field is absent. These two points explain the robustness of the quantization of $\rho_{xy}$ against disorder. It is important to note we do not require any disorder; it plays no essential role in our theory of $\rho_{xy}$ in the IQHE, in keeping with the experimental trend of still observing the IQHE in ever cleaner samples \cite{Pfeiffer2003a}. Also electrons confined to closed interior orbits in strong magnetic fields, while undoubtedly present, play no significant role in our theory.

Certain values of the Fermi energy are of special importance and directly related to integer and half-integer structures: Whenever the magnetic field is such that the Fermi energy falls between two Landau levels, the LDOS is highly suppressed and we obtain an integer filling factor $\nu$ and recover the IQHE with quantized conductivities $\sigma_{xy,\nu}=\frac{\nu e^2}{h}$.  However there is a second interesting case, related to the substructure present in the LDOS. For the filling factors $\nu=\frac{7}{2},\frac{5}{2}$, the LDOS has a zero in the middle of the Landau level, whose position only very weakly depends on the electric field value. The filling factors $\frac{7}{2}$ and $\frac{5}{2}$ are located in the middle of the second Landau level. Both filling factors are separated in energy due to the  Zeeman splitting caused by the electron spin.

For higher electric fields (caused by higher currents) we observe (Figure 3b) the widening of the zeroes of the LDOS into gaps at $\nu=\frac{5}{2}$ and $\nu=\frac{7}{2}$. At the same time the LDOS in between the neighbouring spin-split Landau levels is enhanced and the $\nu=3$ integer gap is no longer present. The theory (Fig.~4 and Fig.~3c) predicts the positions and widening of the gaps at the level centers, and the coalescence of their shoulders, in full agreement with the experiments. Thus the new features in the experiments can be seen as 
of the LDOS gaps \textit{in the center of the second Landau level} and provide conclusive evidence for the validity of the injection model of the quantum Hall effect.

\section*{Acknowledgements}

We thank Y.R.~Li, P.T.~Lin, T.L.~Lin, Y.S.~Tseng, C.K.~Yang, and M.R.~Yeh for experimental assistance. The work undertaken in Taiwan was funded by the NSC, Taiwan. We appreciate stimulating discussions with P.~Kramer and M.~Kleber. T.K.\ is supported by the Emmy-Noether program of the DFG. We would also like to thank the National Nanotechnology Infrastructure Network (NNIN), USA, for computing resources.\\[1ex]
C.F.H.\ and L.H.L.\ initiated the experimental part of the project; C.T.L.\ coordinates the low-temperature measurement facilities; K.Y.C.\ conducted the experiments with C.T.L.; J.Y.W.\ fabricated the sample and S.D.L.\ grew the GaAs/AlGaAs heterostructure. T.K., E.J.H., and R.E.P.\ worked on the theoretical description and modelling.\\[1ex]
Correspondences should be addressed to T.K. (tobias.kramer@physik.uni-regensburg.de),\\ or E.J.H. (heller@physics.harvard.edu), 
or C.-T.L. (ctliang@phys.ntu.edu.tw).

\appendix

\section{The Hall potential in the injection region}

The distribution of the Hall potential close to the current injecting Ohmic contacts is crucial for deriving the current-voltage relation. The analytic solution of Laplace's equation with the boundary condition of a given voltage difference $V_{SD}$ between the two Ohmic contacts and no current leaving the sample along the long sides  \cite{Wick1954a,Thompson1969a,Kawaji1978a,Rendell1981a} leads to a highly non-uniform potential and current distribution, shown in Figure~1. The electric field lines close to two opposite corners of the device follow quarter-circles, whereas in the middle of the device a smaller homogeneous electric field with parallel field lines prevails. Several experiments have imaged either the Hall potential directly  \cite{Knott1995a,Ahlswede2001a} or the resulting hot spots in opposite corners of the device  \cite{Kawano1999a,Ikushima2007a} and thus established the validity of the classical Hall field calculation with its high electric field values at two corners even in the quantum Hall regime. The detailed study of the Hall potential by scanning force microscopy (SFM) in Ref.~\cite{Ahlswede2001a}, Fig.~3, has shown that very close to the injection region ($<0.5\;\mu$m), the Hall profile retains a universal shape, not affected by screening and (in-)compressible stripes which can be present further away 
($>6\;\mu$m) from the injection point, where the amplification of the electric field is reduced.

\subsection*{The solution of Laplace's equation (general Hall angle)}

We follow Ref.~\cite{Rendell1981a} for the calculation of the electric field distribution, the current density, and the potential. 
We solve Laplace's equation subject to the boundary conditions that $V(0,y)=V_{SD}$ and $V(L,y)=0$, in connection with the magnetoresistance equations for the current density $\mathbf{j}$ in the presence of a magnetic field $\boldsymbol{{\cal B}}=(0,0,{\cal B})$, and an electric field $\boldsymbol{{\cal E}}=({\cal E}_x,{\cal E}_y)$ (see  \cite{Beer1963a}, Eq.~(9.1)):
\begin{equation}
\boldsymbol{{\cal E}}+\rho_{yx} \mathbf{j}\times\frac{\boldsymbol{{\cal B}}}{{\cal B}}=\rho_{xx}\mathbf{j}.
\end{equation}
We impose the boundary condition $j_y(x,0)=j_y(x,W)=0$ and thus the electric field components at the point $(x,y)$ is given by
\begin{equation}\label{eq:laplace_efield}
{\cal E}_x(x,y)+\rmi\;{\cal E}_y(x,y)=-\frac{V_{SD}}{W}\exp\left[\frac{4\theta_H}{\pi} 
\sum_{n=1,3,5,\ldots}\frac{\sinh (n \pi [y-W/2+\rmi\;x]/L)}{n\cosh(n\pi W/(2L))}\right],
\end{equation}
where  $\theta_H$ denotes the Hall angle, $L$ the length along the $x$-direction, $W$ the width of the Hall bar (along the $y$-direction), and $V_{SD}$ the voltage difference between the two Ohmic contacts. The potential is obtained by integrating the $x$-component of the electric field along $x$
\begin{equation}\label{eq:laplace_potential}
V(x,y)=-e \int_{L}^{x} \rmd x' {\cal E}_x(x',y).
\end{equation}
The classical current density $(j_x,j_y)$ at point $(x,y)$ is given by
\begin{eqnarray}\label{eq:laplace_current}
j_x(x,y)&=&\frac{{\cal E}_x(x,y)+\lambda {\cal E}_y(x,y)}{\rho_{xx}(1+\lambda^2)},\\
j_y(x,y)&=&\frac{{\cal E}_y(x,y)-\lambda {\cal E}_x(x,y)}{\rho_{xx}(1+\lambda^2)},
\end{eqnarray}
where 
\begin{equation}
\lambda=\tan\theta_H=\frac{{\cal E}_y}{{\cal E}_x}=\frac{\rho_{yx}}{\rho_{xx}}=\frac{\sigma_{xy}}{\sigma_{xx}}.
\end{equation}

\subsection*{The solution of the Laplace and Schr\"odinger equation (Hall angle close to 90$^\circ$)}

In the present experiment a Hall bar with a large length to width aspect ratio $L/W\gg1$ is used and thus the potential obtained in Eq.~(\ref{eq:laplace_potential}) can be considerably simplified in two regions:
in the middle of the device at  $x=L/2$, where it becomes the solution of a parallel-plate capacitor, with plates parallel to the $x$-axis at $y=0,W$ and a potential difference of $V_{SD}$:
\begin{equation}
V_{\text{middle}}(x,y)=\frac{V_{SD}}{W} y.
\end{equation}
The resulting electric field is constant and directed in the $y$-direction:
\begin{equation}
\boldsymbol{{\cal E}}_{\text{middle}}=\frac{V_{SD}}{W}\mathbf{e}_y.
\end{equation}
Another most important simplification happens in the hot-spot corner of the device. There, the solution is given by the electrostatic potential of a rectangular corner, where the two sides have a potential difference $V_{SD}$:
\begin{equation}
V_{\text{corner}}(x,y)=\frac{2 V_{SD}}{\pi} \arctan(y/x), \quad x,y > 0,
\end{equation}
or in cylindrical coordinates $(x,y)=r(\cos\theta,\sin\theta)$ with respect to the corner
\begin{equation}
V_{\text{corner}}(r,\theta)=\frac{2 V_{SD}}{\pi}  \; \theta, \quad 0\le\theta\le\pi/2.
\end{equation}
The electrical field points along the unit vector $\mathbf{e}_\theta$ and is given by
\begin{equation}
\boldsymbol{{\cal E}}_{\text{corner}}(r,\theta)=\frac{2}{\pi} \frac{V_{SD}}{r} \mathbf{e}_\theta.
\end{equation}
The corner and the middle of the device are covered by a single conformal mapping. The existence of a conformal map allows to introduce a set of local orthogonal coordinates. In the middle of the device, the cartesian coordinates $x,y$ form an orthogonal grid, whereas near the corner $r,\theta$ are orthogonal. For convenience, we choose the symmetric gauge with ${\cal A}={\cal B}(-y/2,x/2,0)$. In the middle of the device the Hamiltonian is given by
\begin{equation}
H_{\text{middle}}=
\frac{p_x^2+p_y^2}{2m}
+\frac{1}{2}m\omega_L^2(x^2+y^2)
+\omega_L (p_x y-p_y x)
+\frac{V_{SD}}{W}y,\quad \omega_L=\frac{e {\cal B}}{2m}.
\end{equation}
Introducing cylindrical coordinates $(x,y)=r(\cos\theta,\sin\theta)$ and the angular momentum operator ${\cal L}_z$, we obtain
\begin{equation}
H_{\text{middle}}=
\frac{p^2}{2m}
+\frac{1}{2}m\omega_L^2 r^2
+\omega_L {\cal L}_z
+\frac{V_{SD}}{W}y.
\end{equation}
Similarly, in the corner
\begin{eqnarray}
H_{\text{corner}}&=&\frac{p_x^2+p_y^2}{2m}+\frac{1}{2}m\omega_L^2(x^2+y^2)+\omega_L {\cal L}_z
+\frac{2}{\pi} V_{SD} \arctan(y/x)\label{eq:Hcorner}\\
&=&\frac{p^2}{2m}+\frac{1}{2}m\omega_L^2 r^2+\omega_L {\cal L}_z+\frac{2}{\pi} V_{SD} \;\theta.
\end{eqnarray}
It is important to realize that the potential is in both cases just proportional to one of the orthogonal coordinates. Next, we discuss the local density of states (LDOS) close to the corner.

\section{The density of states in crossed electric and magnetic fields}\label{sec:LDOS}

For the uniform field case with Hamiltonian
\begin{equation}
H_{\text{middle}}=\frac{p_x^2+p_y^2}{2m}
+\frac{1}{2}m\omega_L^2(x^2+y^2)
+\omega_L (p_x y-p_y x)
+\frac{V_{SD}}{W} y,
\end{equation}
the LDOS is derived in  \cite{Kramer2003a}. The LDOS (without spin) in crossed electric and magnetic fields is given by
\begin{equation}
\label{eq:LDOS}
n_{{\cal E}\times{\cal B}}(\mathbf{r};E) = \sum_{k=0}^\infty \frac1{2^{k+1} k! \pi^{3/2}l^2\Gamma} \, \rme^{-E_k^2/\Gamma^2} \, {\left[{\rm H}_k\left(E_k/\Gamma\right)\right]}^2.
\end{equation}
Here, the level width parameter
\begin{equation}
\Gamma = e {\cal E}_\perp l
\end{equation}
is related to the magnetic length $l = \sqrt{\hbar/(e{\cal B})}$, the electric field ${\cal E}_\perp=V_{SD}/W$, and $E_k$ denotes the effective energy shift for the $k$th level:
\begin{equation}
E_k=E-\Gamma^2/(4\hbar\omega_L)-(2k+1)\hbar\omega_L,\quad \omega_L=\frac{e{\cal B}}{2 m}.
\end{equation}
The electron spin leads to a shift of the LDOS and we obtain the LDOS in the presence of spin,
\begin{equation}\label{eq:LDOSspin}
n^{\uparrow\downarrow}_{{\cal E}\times{\cal B}}(\mathbf{r};E) =
n_{{\cal E}\times{\cal B}}\left(\mathbf{r};E - \frac{1}{2}g^*\hbar\omega_L\right)+
n_{{\cal E}\times{\cal B}}\left(\mathbf{r};E + \frac{1}{2}g^*\hbar\omega_L\right).
\end{equation}
For the Hamiltonian
\begin{equation}
H_{\text{corner}}=\frac{p_x^2+p_y^2}{2m}
+\frac{1}{2}m\omega_L^2(x^2+y^2)
+\omega_L (p_x y-p_y x)+\frac{2 V_{SD}}{\pi}  \arctan(y/x),
\end{equation}
we have to perform a numerical calculation of the LDOS by tracking the autocorrelation function of a wavepacket which is propagated using Fast Fourier Transforms  \cite{Kramer2008a}. The resulting LDOS is to a very high accuracy (for $V_{SD}=5$~mV better than $10^{-4}$) approximated by eq.~(\ref{eq:LDOS}), provided we set 
\begin{equation}
{\cal E}_\perp=\frac{2 V_{SD}}{\pi\sqrt{x^2+y^2}}.
\end{equation}
For smaller $V_{SD}$ we obtain an even stronger numerical bound. The accuracy is best tested by subtracting the analytical available time-dependent autocorrelation function in a homogeneous electric field from the one obtained by numerical propagation.

For the homogeneous case, the expectation value of the kinematic velocity $\boldsymbol{\Pi}/m=(\mathbf{p}-e \boldsymbol{{\cal A}})/m$ through a flux line along $\mathbf{e}_x$ is given by
\begin{equation}
\langle\psi_{E,\text{middle}}|\frac{\boldsymbol{\Pi}}{m}\delta(\hat{x}-x_s)|\psi_{E,\text{middle}}\rangle=\frac{{\cal E}_\perp}{{\cal B}} \mathbf{e}_x.
\end{equation}
Likewise we obtain in the corner geometry for the flux through a line along $\mathbf{e}_\theta$, 
\begin{equation}
\langle\psi_{E,\text{corner}}|\frac{\boldsymbol{\Pi}}{m}\delta(\hat{r}-r_s)|\psi_{E,\text{corner}}\rangle
=\frac{2 V_{SD}}{\pi\sqrt{x^2+y^2}{\cal B}} \mathbf{e}_r.
\end{equation}

\section{Quantum mechanical Hall conductivity in the injection region}

While the LDOS (and the velocity) depend on the local field gradient, it is crucial to realize that the gaps of the LDOS in the middle of even Landau levels survive the averaging process over a wide range of local field gradients. However, other oscillatory structures in higher Landau levels are affected by the field-average and may disappear. The disappearance of the gaps at small currents (here for $I_{SD}<800$~nA) is caused by the convolution of the LDOS with the Fermi-Dirac distribution at the liquid Helium temperature $T=0.3$~K.

In principle, the electric field suffers from the mathematical singularity at the corner where $r=0$. However, we will assume that the non-uniformity of the contacts will weaken the singularity and thus limit the upper value of the electric field.  An exact model of the environment around the singularity is beyond the scope of the present paper, but the experimentally observed structures allows us to pin down the effective distance from the singularity to $r_s\approx 220$~nm for a magnetic field around 3 Tesla, which leads to a 300-fold amplification of the electric field at the injection region compared to the middle of the device. This distance corresponds to $2-4$ cyclotron diameters. In the present modelling, we distribute incoherent emitters with equal angular spacing on a quarter circle of radius $r_s=220$~nm.

The Hall angle in the used high mobility samples is between $85^\circ-90^\circ$ and the current is flowing almost perpendicular to the electric field vector (see Figure 1). The equipotential lines near the hot spots are radial rays, while the lines of equal electric field are quarter-circles. Equipotential lines of equal angular spacing from the corner are transformed into parallel lines of equal distance in the middle of the Hall bar. In the presence of a magnetic field, the electrons move on these equipotential lines. Thus a point source emitting with uniform intensity along the quarter circle will populate the middle of the Hall bar uniformly. In the classical theory, we can calculate the average velocity of the electrons along a slice across the middle of Hall bar of width $W$ and length $L$ using the classical equations of motions:
\begin{eqnarray}
v_{\text{cl,average}}(x=L/2)
&=&\frac{1}{W}\int_{0}^{W}\rmd y\; v_x(L/2,y)\\
&=&-\frac{1}{e \cal B W}\int_{0}^{W}\rmd y\; \frac{\partial V(L/2,y)}{\partial y}\\
&=&\frac{V(L/2,W)-V(L/2,0)}{e{\cal B} W}.
\end{eqnarray}
We obtain the classical Hall coefficient by considering a uniform density of electrons $N_{\text{cl}}$ across the middle of the device which originates from a uniform, isotropic emission from a quarter circle close to the hot spot. The voltage drop across the device is equal to the source-drain voltage $e V_{SD}=V(L/2,W)-V(L/2,0)$. Thus the current becomes just
\begin{equation}
I_{\text{cl}}=e N_{\text{cl}} W v_{\text{cl,average}}(L/2)
=\frac{e N_{\text{cl}} V_{SD}}{{\cal B}},
\end{equation}
and the Hall resistance is given by
\begin{equation}
R_{\text{cl},xy}=\frac{V_{SD}}{I_{\text{cl}}}=\frac{{\cal B}}{e N_{\text{cl}}}.
\end{equation}
Next, we translate the classical considerations into the corresponding quantum mechanical model. As in the classical model, we consider the position of the hot-spot at the current injecting edge as the electron source for the current through the device. However, this time we study the quantum-mechanical emission and propagation of the electrons. To this end, we model the current injecting hot spot by a quarter-circle with a fixed radial distance $r_s$ from the mathematical singularity at $r=0$. The Hamiltonian in the vicinity of the corner is given by Eq.~(\ref{eq:Hcorner}). We calculate the contribution to the total current of each source at position $\mathbf{r}_s$ along the quarter circle by integrating over the product of the LDOS $n^{\uparrow\downarrow}_{{\cal E}(\mathbf{r}_s) \times {\cal B}}(\mathbf{r}_s;E)$ at the emission point with the expectation value of the kinematic velocity of the emitted particle:
\begin{eqnarray}\label{eq:jsource}
I_{\text{qm}}
&=&e \int_0^{\pi/2}\rmd \theta
\int_{-\infty}^{\infty}\rmd E\,
\frac{n^{\uparrow\downarrow}_{{\cal E}(\mathbf{r}_s) \times {\cal B}}(\mathbf{r}_s;E)\;
\langle \psi_{E}| \frac{\mathbf{e}_\theta\cdot\hat{\boldsymbol{\Pi}}}{m} \delta(\hat{r}-r_s) |\psi_{E}\rangle}
{\rme^{(E-E_F)/(k_B T)}+1}\\
&=&e \int_0^{\pi/2}\rmd \theta \; r_s \frac{2 V_{SD}}{\pi {\cal B} r_s}
\int_{-\infty}^{\infty}\rmd E\,
\frac{n^{\uparrow\downarrow}_{{\cal E}(\mathbf{r}_s) \times {\cal B}}(\mathbf{r}_s;E)
}
{\rme^{(E-E_F)/(k_B T)}+1}\\
&=&\frac{e V_{SD}}{\cal B} 
\int_{-\infty}^{\infty}\rmd E\,
\frac{n^{\uparrow\downarrow}_{{\cal E}(\mathbf{r}_s) \times {\cal B}}(\mathbf{r}_s;E)
}
{\rme^{(E-E_F)/(k_B T)}+1},
\end{eqnarray}
where we used the results for the LDOS and the velocity given in Sect.~\ref{sec:LDOS}. For evaluating the above equation along the quarter-circle, we take the dissipative nature of the hot-spot into account by launching wavepackets with the same initial kinetic energy,
independent of their starting point on the circle. This assumption is supported by two experimental observations:
(i) the hot-spot shows indeed an increased temperature compared to the rest of the device, and 
(ii) the QHE and the unique fingerprint of the modulations within a Landau level are prominently visible even for Hall potentials exceeding the energetic difference between two Landau levels by a factor of 10.

The conformal mapping assures that the isotropic emission from the corner flowing from the quarter circle is transformed into a homogeneous flow across the full width of the device $W$. Also the potential drop along the quarter circle is equal to the potential drop across the middle of the device. The quantum mechanical expression for the Hall resistivity becomes
\begin{equation}
R_{\text{qm},xy}=\frac{V_{SD}}{I_{\text{qm}}}
=\frac{{\cal B}}
{e \int_{-\infty}^{\infty}\rmd E\,
n^{\uparrow\downarrow}_{{\cal E}(\mathbf{r}_s) \times {\cal B}}(\mathbf{r}_s;E)/{(\rme^{(E-E_F)/(k_B T)}+1)}}.
\end{equation}
The local density of states given in Eq.~(\ref{eq:LDOSspin}) depends strongly on the magnetic and electric field values. Between two Landau levels the LDOS is exponentially suppressed, and it is broadenend and modulated within each Landau level with a zero in the center of even Landau levels.

\section{Self-consistent solution for the source-drain-voltage and current}

Now we are in a position to self-consistently solve the equations for given magnetic field ${\cal B}$, temperature $T$, Fermi energy $E_F$, and current $I_{SD}$ (which is fixed in the experiment by a constant-current source). Experimentally, the gate voltage $V_g$ can be used to change the Fermi energy and thus the average electron density. In this experiment $V_g=0$ is fixed, and thus the Fermi-energy is given by the average electron density $N_s$ divided by the average DOS $n^{\uparrow\downarrow}_{\text{av}}=m^*/(\pi\hbar^2)$, yielding $E_F=N_s/n^{\uparrow\downarrow}_{\text{av}}=8.9$~meV. 

To calculate the source-drain-current -- Hall-voltage characteristics of the device, we fix the Hall angle $\theta_H=90^\circ$. Then we proceed in the following way:
\begin{enumerate}
\item Make an initial guess of the source-drain voltage $V_{SD}$ (i.e. by using the classical Hall effect).
\item From Eq.~(\ref{eq:laplace_efield}), obtain the electric Hall field at a distance $r_s$ from the singularity.
\item Using Eq.~(\ref{eq:jsource}), calculate the total current $I_{qm}$.
\item Compare $I_{qm}$ with the experimentally given value $I_{SD}$. If the current is too large, reduce $V_{SD}$, otherwise
increase $V_{SD}$.
\item Repeat these steps until convergence with respect to $|I_{SD}/I_{qm}-1|< 10^{-4}$ is reached.
\end{enumerate}

Certain ranges of the magnetic field are of special importance and directly related to integer and half-integer structures: Whenever the magnetic field is such that the Fermi energy falls between two Landau levels, the LDOS is highly suppressed and we obtain the integrated LDOS
\begin{equation}
N=\int_{-\infty}^{\infty}\rmd E\, \frac{n^{\uparrow\downarrow}_{{\cal E}(r_s) \times {\cal B}}(\mathbf{r}_s;E)}{\rme^{(E-E_F)/(k_B T)}+1}
=\frac{\nu e {\cal B}}{h},
\end{equation}
with filling factor $\nu=1,2,3,\ldots$. This yields the total current
\begin{equation}
I_{qm}(E_F,T)= \frac{\nu e^2}{h} V_{SD}.
\end{equation}
Thus we recover the integer QHE with quantized conductivities $\sigma_{xy,\nu}=\frac{\nu e^2}{h}$.  However there is a second interesting case, related to the substructure present in $n^{\uparrow\downarrow}_{{\cal E}\times{\cal B}}(\mathbf{r}_s;E)$. For the filling factors $\nu=7/2$ and $5/2$, the LDOS has a zero in the middle of the Landau level, whose position only very weakly depends on the electric field value (see Figure 4). The filling factors $7/2$ and $5/2$ are located in the middle of the second Landau level. Both filling factors are separated in energy due to the  Zeeman splitting $E_{z}=g^*/m^*\mu_{\cal B} {\cal B}$ caused by the electron spin with an effective g-factor. 
For higher electric fields (caused by higher currents) we observe an overlap of the neighboring Landau levels and an enhanced DOS between the two Landau levels. However, the zeroes of the DOS around the two filling factors stand out at all field values, since they actually widen at higher fields. The appearance and widening of the gaps at $\nu=7/2$ and $5/2$ and the coalescence of the shoulders of the two levels are clearly visible in the experiment (Figure 3b) and the theoretical calculation (Figure~3c).

\section{Dissipation and Ohmic resistance}

Two main mechanisms lead to additional resistance in the system: the Ohmic contacts into the device (see the discussion of hot-spots above), and the redistribution of the emitted current in the device. From imaging the vicinity of a hot-spot using SFM  \cite{Ahlswede2001a}, Fig.~3, the following picture emerges: At distances closer than 500~nm from the hot-spot the Hall potential becomes independent of slight variations of the magnetic fields, while starting at a distance of several $\mu$m away from the injection region a redistribution of the Hall potential takes place, which depends on the magnetic field. The above experimental findings are in line with our theoretical model of the origin of the Hall plateaus, which predicts a more robust $\rho_{xy}$ which is unaffected by screening, while the current distribution and voltage drop along the sample edges associated with $\rho_{xx}$ are strongly affected by readjustments of the density caused by screening and disorder. The calculation of the screened potentials and (in-)compressible stripes, relevant for determining $\rho_{xx}$ along the middle section of the device, has been performed in Refs.~ \cite{Siddiki2004a,Siddiki2007a}, but in our model these stripes are not the reason for the existence of the Hall plateaus in $\rho_{xy}$.

Since  $\rho_{xx}$ is probed at a considerable distance from the hot-spot, it is more susceptible to disorder, screening, and gate structures and its calculation relies on more details of the sample than the calculation of $\rho_{xy}$:
In the injection region the disorder effects are negligible because of the huge electric field strength imposed by the device geometry. 
The decoupling of the theoretical description of $\rho_{xy}$ from the one of $\rho_{xx}$ is an important consequence of our model: It can explain the generally observed differences between $\rho_{xx}$ and $\partial \rho_{xy}/\partial {\cal B}$. However, some similarity still prevails, which can be seen in Figure 2b, where the lack of  available states right at the center of a Landau level is also reflected in a dip in the longitudinal resistivity $\rho_{xx}$ around ${\cal B}=4.25$~T.

\section{Relation to the breakdown of the IQHE}

The current densities used in the present experiments are much lower than the one reported for the breakdown of the IQHE  ($\sim 1/100$~Am$^{-1}$ here, vs.\ $\sim 1$~Am$^{-1}$ in Ref.~\cite{Kawaji1996a}). In Ref.~\cite{Kawaji1996a} the breakdown is  characterized in terms of the longitudinal $\rho_{xx}$ component of the resistivity between two adjacent Landau levels. 
The breakdown can be also theoretically described within our model of a current-dependent LDOS, see Ref.~\cite{Kramer2005c}.

\providecommand{\url}[1]{#1}

\end{document}